\documentclass[reprint,amsmath,amssymb,aps,pre,showpacs,twocolumn,floatfix]{revtex4}
\usepackage{epsfig}
\usepackage{graphicx}
\usepackage{makeidx}
\usepackage{chemarr}
\usepackage{dcolumn}
\usepackage{bm,multirow}
\usepackage{color}
\usepackage{bm}

\def\Prob{{\mbox{Prob}}}

\def\epsilon{\varepsilon}

\def\beqr{\begin{eqnarray}}
\def\eqnr{\end{eqnarray}\noindent}
\def\beq{\begin{equation}}
\def\bc{\begin{center}}
\def\ec{\end{center}}
\def\eqn{\end{equation}\noindent}
\begin{document}
\title{  Power spectrum of  mass  and activity fluctuations in a   sandpile}
\author{  Avinash Chand Yadav$^{1}$, Ramakrishna Ramaswamy$^{1,2}$, and Deepak Dhar$^{3}$}
\affiliation{$^{1}$School of Physical Sciences, Jawaharlal Nehru University, New Delhi 110 067, India}
\affiliation{$^2$University of Hyderabad, Hyderabad 500 046, India}
\affiliation{$^{3}$Department of Theoretical Physics, Tata Institute of Fundamental Research, Homi Bhabha Road, Mumbai 400 005, India}

\begin{abstract}
{ We consider  a directed abelian sandpile on a strip of size $2\times n$, driven by adding a grain randomly at the left boundary after every $T$ time-steps. We establish the exact equivalence of the problem of mass fluctuations in the steady state and the number of zeroes in the ternary-base representation of the position of  a random walker  on a ring of size $3^n$.  We find that while the fluctuations of mass have a power spectrum that varies as $1/f$ for frequencies in the range $ 3^{-2n} \ll f \ll 1/T$, the activity fluctuations in the same frequency range have a power spectrum that is linear in $f$. }
\end{abstract}

\maketitle

\section{Introduction}

A subject of continuing interest in the study of non-equilibrium systems is the question of whether there is a common underlying mechanism in the very wide variety of systems where $``1/f"$ noise is observed \cite{1,2,3,4,5,6,31,32,33,34,35}. More generally, the spectrum is a power law, $1/f^{\alpha}$ with $\alpha$ near 1. Further, the exponent $\alpha$ can be different for the noise in different physical quantities in the same system. In the past several decades it has been realized that  $1/f^{\alpha}$ spectra  can arise in a number of different contexts, including fractal renewal processes \cite{38}, constrained reversible Markov chains \cite{40}, nonlinear stochastic differential equations \cite{37}, or multiplicative random processes \cite{39}. 

Bak, Tang, and Wiesenfeld (BTW) proposed  that for a large class of driven non-equilibrium systems, the $1/f^{\alpha}$ noise is a manifestation of the long-ranged  temporal correlations of  fluctuations in their self-organized critical (SOC) steady state \cite{7,8}, and introduced simple driven automaton model of sandpiles that reached a state characterized by power--law time and space correlations. Criticality does not necessarily imply that the exponent $\alpha$ of the power--spectrum is  approximately $1$  \cite{9,10,11}. 

Studies  of noise in self-organized critical systems have generally used the sandpile models as a paradigm. Most of these have been made in the slow driving limit, when the time-interval $T$  between consecutive particle addition events is much greater than the typical duration of an avalanche. In this limit, for frequencies $1 \gg f \gg 1/T$, the time-correlations in the noise-signal can be expressed in terms of the critical exponents that characterize the scaling of the probability distribution of a single avalanche, and the exponent $\alpha$ is also expressible in terms of these exponents \cite{laurson}.    Numerical results  indicate that the activity in the BTW model  exhibits a non-trivial $1/f^{\alpha}$ power spectrum, with the exponent $\alpha$ being between $0$ and $2$ \cite{laurson}.  Some variants of the BTW model also exhibit a nontrivial $1/f$ power spectrum \cite{13, Redig, DD, HJ}.  Numerical simulations of the avalanche spectrum of  the Manna model in 1, 2, and 3 dimensions gave exponents that ranged from 1.44 to 1.9 \cite{laurson}.  The $1/f$ noise in the fluctuations of mass was first seen in a sandpile model with threshold dissipation  (the sand is not  conserved in topplings when the initial height is too large) by Ali \cite{ali}. 

For $1$-dimensional models, scaling arguments and numerical studies have shown  that in the single avalanche regime, the power-spectrum of the fluctuations in the total activity has an exponent  $\alpha$=1. There is much less work dealing with noise correlations for durations much greater than $T$. These are governed by correlations between different avalanches, which are much harder to determine. Hwa and Kardar \cite{hwa} identified three different regimes of frequencies  or time-scales, corresponding to typical duration of a single-avalanche, interval between of successive particle additions,  and the relaxation time of the medium.
They argued that the effective exponent $\alpha$ takes different values in different regimes. For $f \ll 1/T$, the avalanches are anticorrelated, and  one typically finds that the power spectrum varies as $f^{\alpha'}$, with $\alpha' > 0$. Maslov, Tang and Zhang (MTZ)     \cite{14}  studied a one-dimensional directed model of sandpiles, and showed that $\alpha$=1 for noise  in the spectrum of fluctuations of mass for $f \ll 1/T$. 

The above fact---and the analytical tractability of 1--dimensional sandpile models---make the study of 1--dimensional  sandpile models especially interesting, and in the present paper we study  the model of sandpiles introduced by  MTZ in more detail. For a sandpile on a $2 \times n$ strip, while the mass fluctuations have a $1/f$ spectrum for a very wide range of frequencies $1  \gg f T \gg  9^{-n}$, the fluctuation of activity in the same range of frequencies show a power spectrum that is proportional to $f$.   We show  that mass fluctuations  in this sandpile can be analyzed in terms of the position of an unbiased  random walker  on a 1-dimensional ring of $3^n$ sites.  The mass of the pile is related to the {\it number of zeroes} in a ternary representation of the position of the walker. While the random walk on a ring is a rather classical problem, the number of zeroes in the binary (or ternary) representation of its positional coordinate is an interesting object that does not seem to have been studied before.  It is obvious  that this has very slowly decaying correlations, and so long as the mean square deviation of the walker varies as a power of the time,  the variance of the change in this number in time $t$ is expected to vary as $\log t$. This in turn implies that spectrum of mass fluctuations has a $1/f$ tail for a very large range of  frequency. Note that this mechanism underlying  $1/f$ spectrum differs from earlier approaches that obtain it as a superposition of multiple Lorentzians coming from a distribution of relaxation times \cite{HJ}.

The plan of this paper is as follows. In Section II we briefly describe the sandpile model on a $2 \times n$ strip. We then present the details of the ternary base representation, and  the mapping to the random walk on a ring. In Section III we calculate the time correlation of mass fluctuations in the sandpile using the new ternary representation and in Section IV we compute the local activity, its mean, and the asymptotic velocity of an avalanche wave. In Section V we calculate the mean total activity and present  numerical results for the  power spectrum of its fluctuations.  The activity fluctuations scale as the time-derivative of mass and thus have a power spectrum proportional to $f$. Section VI  contains a summary of our results.

\section{Definition of the model and its relation to a random walk}

The model studied earlier by MTZ is a directed sandpile \cite{15} on a strip of dimension $2\times n$, shown in Fig.~{\ref{fig1}. An integer variable $z(x,y)$ is assigned to each site $(x,y)$ of the lattice: this is the ``number of grains of sand" (or height) at site $(x,y)$. The critical height is $z_{c} = 1$ and if $z(x,y) > z_{c}$, that site topples and transfers two grains,  one to its right neighbor and one to the other site in the same column. One grain leaves the pile  for each toppling that occurs at the rightmost column, $x=n$.

The sandpile is driven periodically with a  grain of sand  added after every $T$ time steps 
at one of the two sites in the first column, selected  randomly.  If as a result of this process the site becomes unstable,  we relax the configuration by toppling all unstable sites in parallel, until all sites are stable. Further, if $T > 2n$ the avalanche activity will have died out before the next particle  is added \cite{footer}.

As noted by MTZ \cite{14} the number of recurrent configurations in this model is $N$ = $3^{n}$. 
In a recurrent configuration, each column $r$ can have the configuration $(\begin{smallmatrix} 1 \\ 0 \end{smallmatrix}), (\begin{smallmatrix} 1 \\ 1 \end{smallmatrix})$ or $(\begin{smallmatrix} 0 \\ 1 \end{smallmatrix})$ , independent of the status of other columns.  Define an integer $\eta_j$ which takes values +1, 0 or $-1$, if the configuration in the $j$th column is $(\begin{smallmatrix} 1 \\ 0 \end{smallmatrix}), (\begin{smallmatrix} 1 \\ 1 \end{smallmatrix})$ or $(\begin{smallmatrix} 0 \\ 1 \end{smallmatrix})$ respectively. Then  to each of the $3^n$ recurrent configurations, we can attach a unique label $j$  given by
\beq
j = \sum_{r=1}^{n}\eta_{r}3^{r-1} + c.
\eqn    
with $c = (3^{n}-1)/{2}$, so that 0 $ \leq j \leq  3^{n} -1$.   Rows are counted from left to right as usual, and it is convenient to adopt the convention that the ternary representation of the integer $j$, namely $\eta_{1}\eta_{2}\ldots\eta_{n}$ is also read left to right. Note that then  the configuration with the label $10\bar{1}0$ corresponds to the integer $j= 1\cdot3^0 + 0\cdot3^1 -1\cdot3^{2} +0\cdot3^{3}+ c$, with $c$ =40 and not $0\cdot3^0 -1\cdot3^1 +0\cdot3^2 +1\cdot3^3 +c $.

\begin{figure}[t]
  \centering
  \scalebox{0.5}{\includegraphics{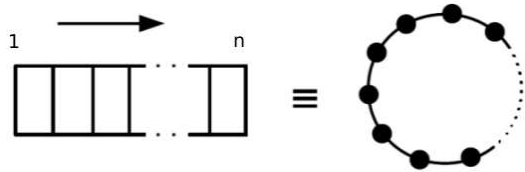}}
  \caption{The directed sandpile on a  $2\times n$ strip (left) is equivalent to a random walk on a ring with $N = 3^n$ sites (right).}
\label{fig1}
\end{figure}
\begin{figure}[t]
  \scalebox{0.65}{\includegraphics{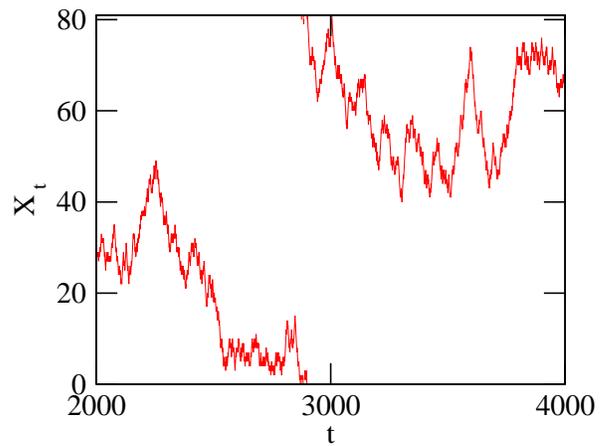}}
  \caption{(Colour online) Typical evolution of a sandpile with $n = 4$, seen as a random walk on a ring of size 81.}
\label{fig2}
\end{figure}

Let $A_{1,1}$ and $A_{1,2}$ denote the operators corresponding to adding a particle at site $(1,1)$ and $(1,2)$ respectively, and relaxing.  These satisfy the abelian property  $[A_{1,1} , A_{1,2}]=0$.  Let the height configuration represented by the integer $j$ be denoted by $\vert j\rangle$. Then we find that  
\beq
A_{1,1} \vert j\rangle \to \vert j +1\rangle;~~~~~  A_{1,2} \vert j\rangle \to \vert j -1\rangle
\eqn
where the addition is understood to be mod $3^n$, so that $|3^n\rangle \equiv |0\rangle$.

The proof is quite straightforward.  For any unstable configuration  with heights 
$\{z(x,y)\}$, we define the variable $ F( \{z(x,y)\})$ by
\begin{equation}
F( \{z(x,y)\}) = \sum_{x=1}^n   3^{x-1} [z(x,1) - z(x,2)].
\end{equation}
It is easy to see that by adding a particle at $(1,1)$ $F$ increases by $1$, while by adding at $(1,2)$  it decreases by $1$. A toppling at any site $(x,y)$, with $x \neq n$ leaves $F$ unchanged, and a toppling at the rightmost column changes $F$ by $\pm 3^n$.

This can also be seen  from  the abelian algebra of the sandpile \cite{16}, using  the fact that $A_{1,1}$ and $A_{1,2}$ can be considered as generators 
of the Abelian group, with the relations \cite{14}
\begin{equation}
A_{1,1} A_{1,2} = ~~~I ~~ =~~A_{1,1}^N ~~= ~~A_{1,2}^N
\end{equation}

Representing the $N$ recurrent configurations of the pile as sites on a ring (see Fig.~\ref{fig1}), stochastic addition of particles to the pile at the left end gives a random walk: at each time step, the walker at site $j$ has equal probability to take a step to the left or the right. We will denote the random walk corresponding to a particular realization of the random evolution of the pile by $W$, and the position of the walker at time $t$ by $X_t$. 
 A typical realization of such a random walk is shown in Fig.~\ref{fig2} for a lattice of size $n = 4$.

\section{The autocorrelation function of mass fluctuations}

The total number of grains in the pile at time $t$ is the mass, denoted by $m(t)$. 
The autocorrelation function $C(\tau)$  of the  fluctuations in mass in the steady state of the pile is
\beq
C(\tau) = \langle m(t) m(t +\tau) \rangle - \langle m(t) \rangle ^2
\eqn
where the angular brackets denote averaging over the steady state.

It is easy to see that $\langle m(t)\rangle = 4n/3$. Let $m_j(t)$ denote the mass of particles in the column $j$. Since in the steady state, the  masses $m_j$ in different columns  are independent random variables,  the equal time mass correlation is 
\beq
C(0) = \sum_{j} \left[\langle m_{j}^2\rangle -   \langle m_{j} \rangle ^2\right]= \frac{2}{9}n.
\eqn

\begin{figure}[t]
  \scalebox{0.65}{\includegraphics{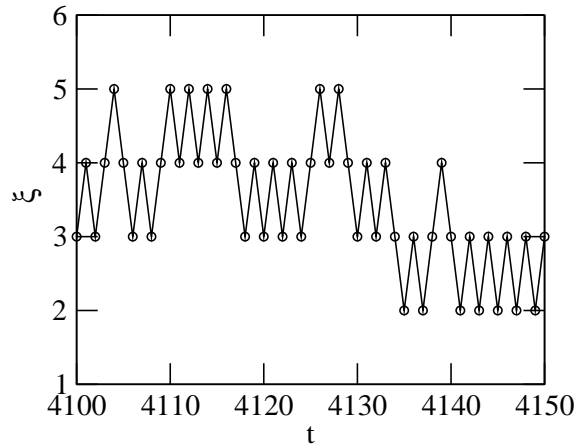}}
  \caption{ Part of the time-series of mass fluctuations $\xi$ for a system of size $n$ = 6.}
\label{fig3}
\end{figure}
\begin{figure}
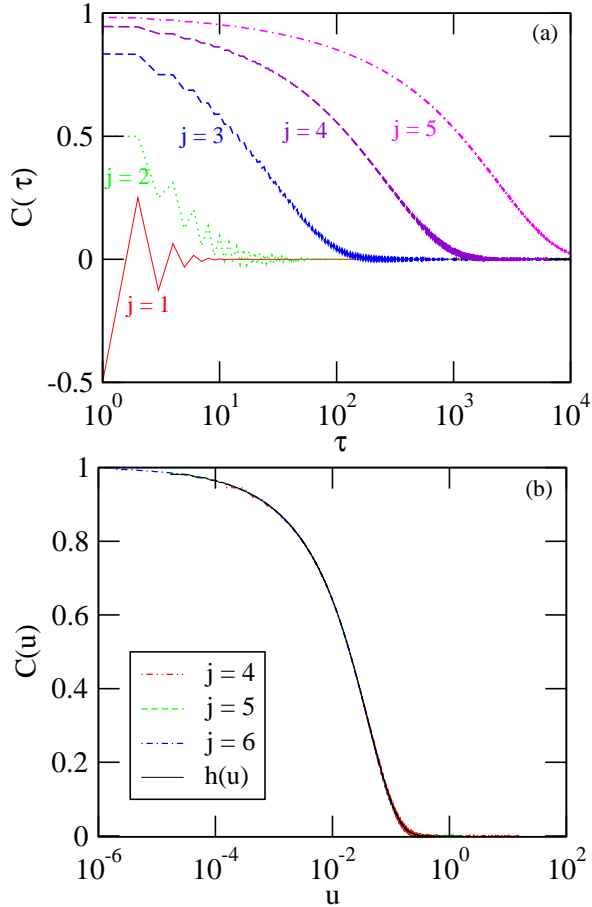

  \scalebox{0.65}{\includegraphics{fig4a.eps}}
  \scalebox{0.65}{\includegraphics{fig4b.eps}}
\caption{(Colour online) (a) Autocorrelation function $C_j(\tau)$ of time series
  $\xi_{j}$ for different values of $j$ for system size $n = 6$. The
  discrete data points are joined by lines as an aid to the eye;  the theoretical curves [Eq.~(\ref{eq12})] are indistinguishable from the numerical results. (b)
  Data collapse of $C_j(\tau)$ for different $j$ onto a single curve with scaling variable $u =  9^{-j} \tau$. }

\label{fig4}
\end{figure}
\begin{figure}[t]
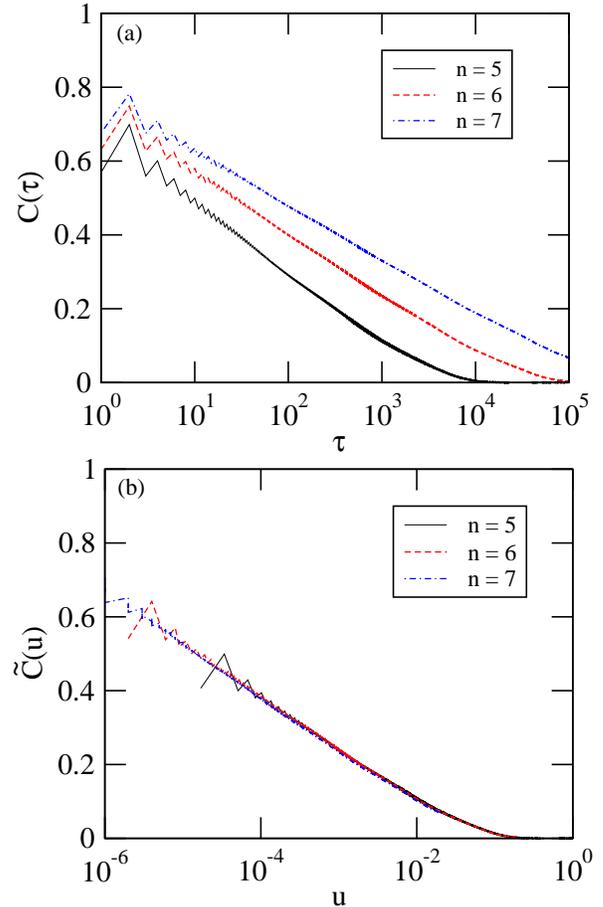

  \scalebox{0.65}{\includegraphics{fig5a.eps}}
  \scalebox{0.65}{\includegraphics{fig5b.eps}}
  \caption{(Colour online) (a) Plot of the autocorrelation  function of mass fluctuations for different system sizes $n$. (b) Collapse onto a single scaling function  when  the scaling variable $u = 9^{-n} \tau$ is used.}
\label{figgc}
\end{figure}
\begin{figure}[t]
  \scalebox{0.65}{\includegraphics{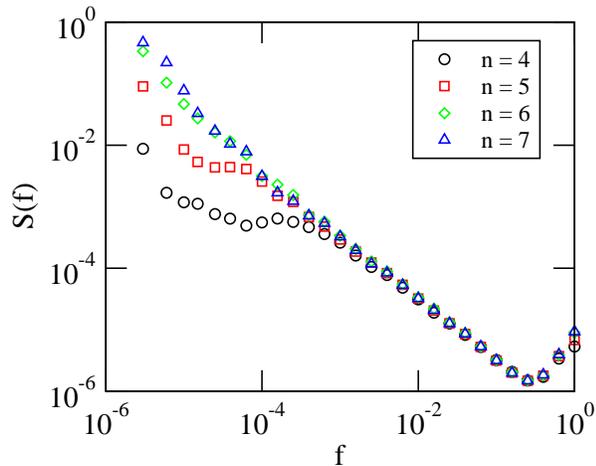}}
  \caption{(Colour online) Power spectrum of  mass fluctuations for different $n$ showing the 1/$f$ dependence.}
\label{figps}
\end{figure}

We define a variable $\xi_{j}$ corresponding to the $j$th bit of the position $X_{t}$ of the walker by  $\xi_{j}(X_{t}) = 1 $ if  $\eta_{j} = 0,$ and  $\xi_{j}(X_{t}) = 0$  otherwise. 
Then the mass $m(t)$ of the pile is
\beq
m(t) = n + \sum_{j=1}^{n} \xi_{j}(X_{t}). 
\eqn

The function $\xi_{j}(X)$ is a (non-random) periodic function of its argument $X$ of period $N_j = 3^{j}$.
For example, $\xi_{1}(X)$ is 0,1 or $-1$, for $X$= 0,1,2~(mod{~3}) respectively.
Expressing  $\xi_{j}(X)$ in terms of its Fourier series expansion
\begin{equation}
\xi_{j}(X) = \sum_{k} \tilde{\xi_{j}}(k) \exp{(ikX)},
\eqn
where $k = {2\pi r}/{N_{j}}$ and $r = 0, 1, \ldots, N_{j}-1$, the coefficients are given by  
\beqr
\tilde{\xi_{j}}(k)  &=& \frac{1}{N_{j}}[1 + e^{-ik} + e^{-i2k} + \dots + e^{i(N_{j-1}-1})k]\nonumber \\
                  &=& \frac{1}{N_{j}} e^{-i(N_{j-1})\frac{k}{2}}[\frac{\sin{(\frac{N_{j-1}k}{2})}}{\sin{(\frac{k}{2})}}].
\label{eq10}
\eqnr

Consider the auto-correlation function $f_j(\tau)$ of the random variables $\xi_{j}$, defined as 
\begin{eqnarray}
f_{j}(\tau) &=&  \langle \xi_{j}(X_{t}) \xi_{j}(X_{t+\tau}) \rangle\nonumber \\
               &=& \sum_{k}\sum_{k'} \tilde{\xi}_{j}(k) \tilde{\xi}_{j}(k') \langle e^{ikX_{t} + ik' X_{t+\tau}}\rangle.
\eqnr
By translational invariance along the ring, this expectation value is zero unless $k+k'=0$. Also, $\langle e^{ikX_{t} - ik X_{t +\tau}}\rangle$ is the characteristic function of the displacement $X(t +\tau) -X(t)$ of a simple random walk $W$ and this is easily calculated, giving 
\begin{eqnarray}
f_{j}(\tau) =\sum_{k} \vert \tilde{\xi}_{j}(k) \vert ^{2} [\cos{k}]^{|\tau|}. 
\eqnr
Using Eq. (\ref{eq10}), this can be evaluated, and the result is  
\begin{equation}
f_{j}(\tau) = \sum_{r} [ \frac{\sin{(\frac{N_{j-1}k_r}{2})}}{N_j \sin{(\frac{k_r}{2})}}] ^{2} [\cos{k_r}]^{|\tau|},
\label{eq12}
\eqn
where $k_r = {2\pi r}/{N_{j}}$, with $r = 0, 1, \ldots, N_{j}-1$.  For $k_r \ll 1$, we can write $[\cos k_r]^{|\tau|} \approx \exp( - k_r^2 |\tau|/2)$. Also, $N_{j-1} k_r$ only takes values $0, 2\pi/3, 4 \pi /3$.  

The $r$=0 term is the disconnected part of the correlation function, and this gives 
a finite contribution, 1/9.  When  $r$ is a nonzero multiple of 3 the numerator vanishes, and 
we have
\begin{equation}
f_j(\tau) \approx h(|\tau|/9^j) + \frac{1}{9},
\eqn
where
\begin{equation}
h(u) = {\sum_{r}}^{\prime}  \frac{3}{4 \pi^2 r^2}  \exp ( - 2 \pi^2 r^2 u).
\label{eq15}
\eqn
The prime denotes that $r$ is not a multiple of 3. As a simple check one can verify  that this 
gives  $h(0)$ = 2/9.

Since $\xi_{j}$ and $\xi_{j'}$ have different periods, these time series are uncorrelated, and 
in the steady state the average of the product $\sum_{j \neq j'} \langle \xi_{j}(t) \xi_{j'}(t') \rangle$ reduces to a product of averages of the individual terms, and thus 
\beq
\sum_{j \neq j'} \langle \delta \xi_{j}(t) \delta \xi_{j'}(t') \rangle = 0.
\eqn
Therefore the connected part of the correlation function can be reduced to  
\begin{eqnarray}
C(\tau) = \sum_{j=1}^{n} C_{j}(\tau) \approx  \sum_{j=1}^{n}  h(9^{-n} |\tau |).
\label{eq17}
\eqnr
In the range $ 9^s < \tau < 9^{s+1}$ the first $s$ terms in the above summation are nearly zero, and each of the remaining terms is nearly equal to $h(0)$. Thus we have   
\begin{equation}
C(\tau) \approx  (2/9) [n - \log_9 {|\tau|}].
\label{eq25}
\eqn
Since the global correlation function varies as $-\log{\tau}$, the resulting power spectrum $S(f)$ will therefore show an asymptotic $1/f^{\alpha}$ dependence with $\alpha =1$, for $1\gg f \gg 9^{-n}$.

Results are shown in Fig. \ref{fig3} for the time--series of mass fluctuations $\xi$. In Fig. \ref{fig4}, we have shown the observed correlation functions $C_j(\tau)$ for different values of $j$. The data was obtained by a Monte Carlo simulation for system size $n=6$, using a time-series of $\xi_j$ of total length 10$^8$ and sample averaged over 10$^2$ realizations. These are also compared with the theoretical values of these functions  for the corresponding correlation functions $C_j(\tau)$ using Eq.~(\ref{eq12}). 

In Fig. \ref{figgc}, we have also shown the  auto-correlation function of fluctuations of the total mass for different values of $n$. We find that data for different $n$ collapse onto a single scaling curve when the scaling variable  $u = \tau/9^n$ is used.   The power spectrum of these mass fluctuations is shown in Fig. \ref{figps} for different system sizes. 

\section{Fluctuations of the local activity}

Define the local activity $a({\bf x},t)$ of site ${\bf x} = (x,y)$ at time $t$ as
1 if the site ${\bf x}$ topples at time $t$ and 0 otherwise.
Addition of a particle at the left edge results in an avalanche wave, the evolution of which  can be graphically depicted by drawing a directed arrow  along a bond if a particle was transferred along it during the avalanche. A typical such wave is shown in Fig.~\ref{fig5}. The avalanche activity can branch, and these branches may rejoin. The maximum time that a wave can take to propagate through the lattice is clearly $2 n$.  
\begin{figure}[b]
  \centering
  \scalebox{0.6}{\includegraphics{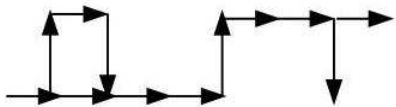}}
  \caption{A typical avalanche wave with branches.}
\label{fig5}
\end{figure}

\begin{figure}[t]
  \centering
  \scalebox{0.65}{\includegraphics{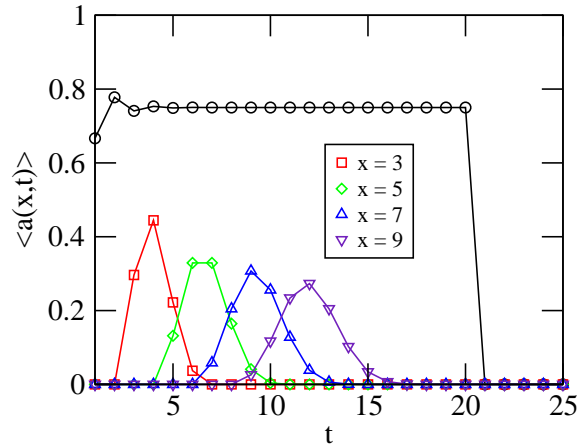}}
  \caption{(Colour online) The mean activity of a column at distance $x$ for a system of size $n = 10$, and $T= 25$. The topmost curve (open circles)  denotes the sum of the local activity in the different columns.}
\label{figseven}
\end{figure}
The  average activity  $\langle a({\bf x},t)\rangle$  at the site $\bf{x}$ at time $t$  in the steady state of the driven pile is  a periodic function with period $T$.  The mean local activity $\langle a({\bf x},t)\rangle$ at site ${\bf x} = (1,1)$ can be computed easily. A particle is added at the top or bottom site with probability $1/2$, and if added at $(1,1)$ and that site becomes active, then the site was originally occupied, the probability of which is $2/3$. The mean local activity  at $(1,1)$ at time $t = 1$ is therefore $\frac{1}{2}\cdot\frac{2}{3} = 1/3$. if the particle is added at the bottom site and it becomes active, then at the next time step it will be active only if the left edge had the configuration  $(\begin{smallmatrix} 1 \\ 1 \end{smallmatrix})$, and this has probability $1/3$. The avalanche wave  begins with  $R_{B}U$  (symbols $R_{B} \equiv _{\longrightarrow}$, $R_{T} \equiv ^{\longrightarrow}$, $U \equiv \uparrow$ and $D \equiv \downarrow$ are used to show arrow representation of avalanche wave) and the mean local activity $\langle a({\bf x},t)\rangle$ at site ${\bf x} = (1,1)$ at  $t = 2$ is $\frac{1}{2}\cdot\frac{1}{3} = 1/6$. The top-bottom symmetry implies that the mean local activity of the bottom site is equal to the mean local activity of the top site. 

In a similar manner, the mean local activity $\langle a({\bf x},t)\rangle$ at site ${\bf x} = (2,1)$ can be calculated. The minimum time taken for the avalanche to reach this site  is $2$. If the wave  is $R_{T}R_{T}$,  it is simple to see that the mean local activity is $\frac{1}{2}\cdot(\frac{2}{3})^{2} = 2/9$. The maximum time taken  is $4$ and the wave is $R_{T}DR_{B}U$ giving activity $\frac{1}{2}\cdot(\frac{1}{3})^{2} = 1/18$. At time $3$ both $R_{B}UR_{T}$ and $R_{B}R_{B}U$ are responsible, giving activity $2\cdot \frac{1}{2}\cdot\frac{1}{3}\cdot\frac{2}{3} = 2/9$.

Generalizing this, one can easily deduce that the local activity of any site 
${\bf x} \equiv (x,y)$ at time step $t$ is ($r = i\mod{x},~ i = t\mod{T}$ and $p = 1/3$)
\beq
\langle a({\bf x},t)\rangle =\begin{cases}
\frac{1}{2} \binom{x}{r}  p^{r}(1-p)^{x-r} & \text{if $x \le {i} \le 2x$;}\\
0 & \text{otherwise.}\end{cases}
\eqn 
The time taken by the avalanche to reach site $n$ is $T_{n} = n + r$ with $n \le T_{n} \le 2n$. 
The asymptotic velocity of the avalanche can be computed, 
\beq
\langle v \rangle  = \lim_{n \to \infty}\langle \frac{n}{ T_{n} } \rangle = 3/4
\eqn	
since the mean local activity follows a binomial distribution, and
$\langle r \rangle = n/3$.  The mean local activity,  shown in Fig.~\ref{figseven}, has a binomial distribution.

\section{The  mean total activity}
Let $\mathit{A}(t)$ denote the total activity in the pile at time $t$,
\begin{equation}
{\mathit{A}}(t) = \sum_x a(x,t).
\eqn
Its mean, averaged over all possible evolutions is ${\mathit{\bar{A}}}(t)$ and the fluctuations are $\delta {\mathit{A}}(t) = {\mathit{A}}(t) - {\mathit{\bar{A}}}(t)$. The power--spectrum of ${\mathit{A}}(t)$ consists of two parts: $\delta$-function peaks at integer multiples of frequency $ \omega_0 = 2 \pi /T$ coming from ${\mathit{\bar{A}}}(t)$ which is a periodic function of time with period $T$, and a continuous part  coming from the fluctuations $\delta {\mathit{A}} (t)$. 

We now discuss the exact calculation of ${\mathit{\bar{A}}}(t)$. This is important for numerical studies since the power--spectrum is dominated by the $\delta$--function peaks, and only after subtracting the periodic part it is possible to obtain a good estimate for the continuous part of the noise--spectrum. 

Let $F_{j}$ be the particle flow out of the $j$th column. There are five options: No flux ($\phi$),  $T, B, TB,$ and $BT$, where $T$ denotes a particle coming out on the top row, $B$ out of bottom row, $TB$ two particles, one from the bottom and then one out of the top row (note that the later event is on the left), and similarly, $BT$.

\begin{figure}[t]
  \centering
  \scalebox{0.65}{\includegraphics{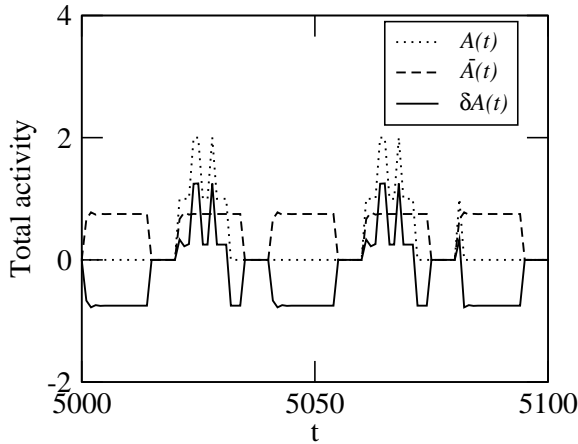}}
  \caption{ Time evolution of total activity $\mathit{A}(t)$ in the pile as a function of time $t$  for a system of size $n = 7$, and with $T =20$. The activity $\mathit{A}(t)$ takes values 0, 1,  or 2. The mean activity $\mathit{\bar{A}}(t)$ and the excess activity $\delta {\mathit{A}}(t)$ are also shown.  }
\label{fig9}
\end{figure}

The time evolution of $F$ as the avalanche moves along the strip can be deduced in a straightforward manner.  Given $F_{j}$ and the configuration of the sandpile at column $|j+1\rangle$,  $F_{j+1}$  we can be determined using the toppling rules:
\begin{itemize}\item
If $F_{j}= T $ and the configuration is $(\begin{smallmatrix} 1 \\ 0 \end{smallmatrix})$, then $F_{j+1} = T.$
\item
If $F_{j}= B $ and the configuration is $(\begin{smallmatrix} 1 \\ 0 \end{smallmatrix})$, then $F_{j+1} = \phi .$
 \item
If $F_{j}= T $ and the configuration is $(\begin{smallmatrix} 1 \\ 1 \end{smallmatrix})$, then $F_{j+1} = BT.$
 \end{itemize}

Define the ket vector $|\alpha \rangle = |\phi \rangle, |T \rangle, |B\rangle, |BT\rangle,$ and $|TB\rangle$ for $\alpha$ = $0, 1, 2, 3, 4$.  Let \Prob($j,\alpha, \tau $) be the probability that the flux at column $j$ is of type $\alpha$, and starts after $\tau$ time steps. The speed change in the avalanche wave can be accounted for by attaching a weight $z$ at each time step, and defining
\beq
|P_{j} \rangle = \sum_{\alpha, \tau}z^{\tau}\Prob(j, \alpha, \tau)|\alpha \rangle.
\eqn 
The initial condition is 
\begin{eqnarray*}
|P_{0} \rangle  = \frac{1}{2}|T \rangle + \frac{1}{2}|B \rangle
\end{eqnarray*}  
from which    
it  can easily be worked out that
\begin{eqnarray}
|P_{1} \rangle  = \frac{1}{3}|\phi \rangle + \frac{z}{6}[|T \rangle + |B \rangle + |TB \rangle + |BT \rangle].\nonumber
\end{eqnarray}
Since each column can have only one of three configurations,  $(\begin{smallmatrix} 1 \\ 0 \end{smallmatrix}), (\begin{smallmatrix} 0 \\ 1 \end{smallmatrix})$ or $(\begin{smallmatrix} 1 \\ 1 \end{smallmatrix})$, the evolution of the toppling wave as it moves along the strip is governed by
the equation
\beq
|P_{j+1} \rangle  = M|P_{j} \rangle
\eqn  
where, $M = \frac{1}{3}(M_{1}+M_{2}+M_{3})$
and $M_{1}$ is a matrix that specifies the change in $|P_{j}\rangle$ when it encounters $(\begin{smallmatrix} 1 \\ 0 \end{smallmatrix})$. Explicitly,
\[
 M_{1} =
 \begin{pmatrix}
  1 & 0 & 1 & 0 & 0 \\
  0 & z & 0 & 0 & 0 \\
  0 & 0 & 0 & 0 & 0 \\
  0 & 0 & 0 & z & z^{2} \\
  0 & 0 & 0 & 0 & 0 
 \end{pmatrix}
\]
It can be easily verified that $M_{1}|\phi\rangle = |\phi \rangle,  M_{1}|T\rangle = z|T \rangle,  M_{1}|TB\rangle = z^{2}|BT \rangle.$ Similarly,
\[
 M_{2} =
 \begin{pmatrix}
  1 & 1 & 0 & 0 & 0 \\
  0 & 0 & 0 & 0 & 0 \\
  0 & 0 & z & 0 & 0 \\
  0 & 0 & 0 & 0 & 0 \\
  0 & 0 & 0 & z^{2} & z
 \end{pmatrix}
\]

and 

\[
 M_{3} =
 \begin{pmatrix}
  1 & 0 & 0 & 0 & 0 \\
  0 & 0 & 0 & 0 & 0 \\
  0 & 0 & 0 & 0 & 0 \\
  0 & z & 0 & z & 0 \\
  0 & 0 & z & 0 & z
 \end{pmatrix}
\]
Then, the mean activity ${\mathit{\bar{A}}}(t)$ at time $t$ is given by the coefficient of $z^{t}$ in $\sum_{j=1}^{n}\langle \psi|P_{j}\rangle$  where $\langle \psi| = \langle 0, 1, 1, 1+z, 1+z|$ and the Fourier or Laplace transform of ${\mathit{\bar{A}}}(t)$ is
\begin{eqnarray}
{\mathit{\bar{A}}}(z) &=&  \sum_{j=1}^{n}\langle \psi|P_{j}\rangle\nonumber\\
          &=& \sum_{j=1}^{n}\langle \psi|(\frac{M_{1}+M_{2}+M_{3}}{3})^{j}|P_{0}\rangle.
\end{eqnarray} 

\begin{figure}[t]
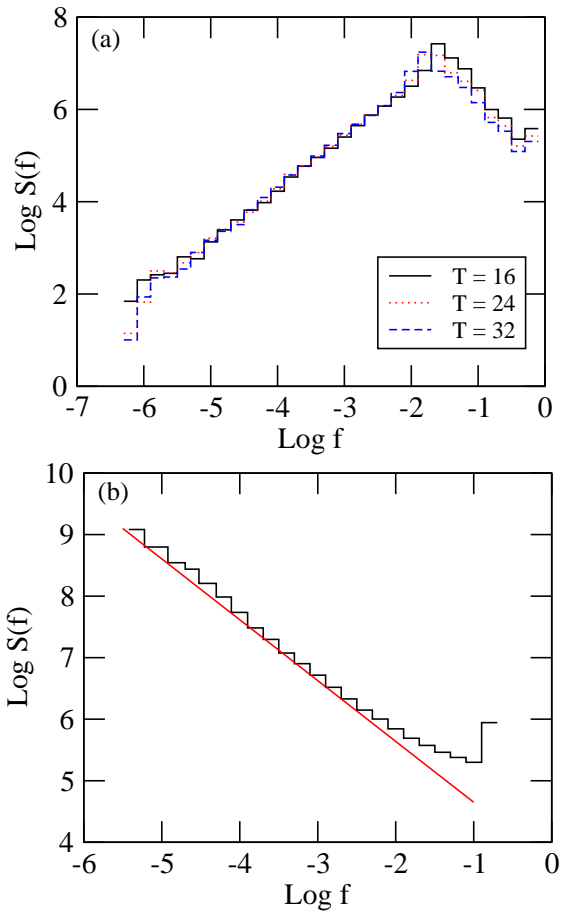

  \centering
  \scalebox{0.65}{\includegraphics{fig10a.eps}}
  \scalebox{0.65}{\includegraphics{fig10b.eps}}
  \caption{(Colour online) The power spectrum of the fluctuations  of excess activity $\delta {\mathit A}(t)$ in a sandpile of size  $n$=7.  (a) The spectrum for different $T$ using a time series of total length $5\times10^6$ showing the initial linear portion as well as the 1/$f$ part, and (b) the spectrum for the case when $T$ is not fixed (a new particle is added when there is no activity)from a time series of total length $10^6$. The power spectrum has been binned using 5 bins per decade of frequency, and a line of slope -1 is shown for comparison.}
\label{fig10}
\end{figure}

The matrix calculation can be simplified by first noting that $|\phi\rangle$ can be dropped from the calculation since it does not contribute. Then from the symmetry of the top and bottom rows the $4 \times 4$ matrix can be reduced to a $2\times 2$ matrix. $|P_{j}\rangle$ is always of the form $\alpha_{j}(|T\rangle + |B\rangle) + \beta_{j}|TB\rangle$
\begin{eqnarray}
 {\mathit{\bar{A}}}(z) =  \sum_{j=1}^{n}\begin{pmatrix} 1 & 1+z \end{pmatrix}\begin{pmatrix} \frac{z}{3} & 0  \\ \frac{z}{3} & \frac{z(z+2)}{3} \end{pmatrix}^{j}\begin{pmatrix} 1 \\ 0 \end{pmatrix}\nonumber\\ 
{\mathit{\bar{A}}}(z) =  \sum_{j=1}^{n}\begin{pmatrix} 1 & 1+z \end{pmatrix}\begin{pmatrix} \frac{z}{3} & 0  \\ \frac{z}{3} & \frac{2z}{3}+\frac{z^{2}}{3} \end{pmatrix}^{j}\begin{pmatrix} 1 \\ 0 \end{pmatrix}
\end{eqnarray} 
and this tridiagonal matrix is easy to evaluate. We get ${\mathit{\bar{A}}}(z) = 2z/{3}+ {7z^{2}}/{9} + \dots $. The mean total activity at time step $k$ can be written as
\begin{eqnarray}
1 - (1/3) + (1/3^{2}) \dots + (1/3^{k})
=\frac{3}{4} [ 1 - (-\frac{1}{3})^{k+1}]
\end{eqnarray}\noindent
which goes to 3/4 as $k \to \infty$.

Shown in Fig.~\ref{fig9} is a typical realization of the total activity ${\mathit A}(t)$ for $n= 7$.  The corresponding power spectrum is  shown in  Fig.~\ref{fig10}. We see that there is a weak dependence of the spectrum on $T$, the period of addition of particles,  shown in Fig.~\ref{fig10}(a) and the power spectrum of these signals exhibit asymptotic  $f$ behavior for $f \ll 1/T$.  This behavior  may be understood as follows: the fluctuations of activity scale the same way as the fluctuations of number of particles leaving the system. When this  number is non-zero, the corresponding avalanche spans the system, and the total activity is also large. When the avalanche wave does not span the system, the activity is less, and the outflux of particles is zero. Hence, we expect that scaling properties of time-correlations of fluctuations of total activity to be similar to that of fluctuations of outflux. But the time series of the latter is obtained by taking successive differences of time-series for fluctuations of mass.  Then, the power spectrum for the fluctuations of total activity
is related to that of mass by an additional factor $f^2$.

Interestingly, if the time interval between additions of  grains is not fixed and a new grain is added as soon as the avalanche generated by the previous grain has stopped, the behavior of activity correlations changes.  In Fig.~\ref{fig10}(b), we have shown the result for power spectrum of total activity fluctuations in such a simulation. We find that the power spectrum shows $1/f$ behavior for several decades of frequency range in this case. It has been noted earlier that the power spectrum for frequencies $f > 1/T$ is not substantially affected by this change \cite{laurson}, but the reason why this extends the range of $1/f$ scaling to much lower frequencies is not understood.

\section{Summary and Discussion}

In this paper we have established an exact equivalence between the total mass--fluctuations  in a directed sandpile on a ladder and the number of zeroes in the ternary representation of a random walker on a ring. We obtained  an exact expression for the time correlation function of mass fluctuations. Local temporal correlations have a characteristic exponential dependence on the  spatial coordinate and from an exact calculation we find that the functional form of the global correlation is logarithmic, resulting in an $1/f$ spectrum for frequencies below a cutoff, $f \le f_{c}$. We also calculated  the mean local activity, the mean total activity and the asymptotic velocity of this avalanche wave was shown to be $ \langle v \rangle  = \frac{3}{4}$. 
For the periodically driven case the power spectrum of total activity fluctuations exhibits $1/f$ dependence for frequencies $f$ satisfying $1 \gg f \gg 1/T$, and a power proportional to $f$ for $1/T \gg f \gg 3^{-2n}$. When  a new particle is added as soon as an avalanche is finished  the power spectrum is $1/f$ over the entire range $ 1 \gg f \gg 3^{-2n}$.

It is also important to note that in this model  the lower cutoff on the range of observed power-law dependence is very small, varying as $\exp(-L)$ where $L$ is the linear extent of the system.  There are related models wherein events relax very very slowly:  for instance in ricepiles where the toppling condition depends on the slope, the mean residence time for a marked grain to leave the pile varies  as $\exp( a L^{d+2})$ for a $d$-dimensional system \cite{pradhan}. It would be interesting therefore to investigate how the lower cutoff  for the $1/f^{\alpha}$ spectrum  depends on system size in other sandpile models, particularly when there are properties that can  have exceedingly long relaxation times.

\section*{ACKNOWLEDGMENTS}
ACY would like to thank the CSIR, India for a Junior Research Fellowship. RR and DD would like to acknowledge the financial support from  the Department of Science and Technology, Government of India through  JC Bose Fellowships, and would also like to thank the ICTP, Trieste for hospitality during the summers of 1990 and 2007 when some version of the present work was started. DD acknowledges a useful discussion with Nick S. Jones on this topic. We thank  S. N.  Majumdar for a critical reading of the manuscript.


\begin{thebibliography}{99}
\bibitem{1} J. B. Johnson, Phys. Rev. {\bf 26}, 71 (1925).
\bibitem{2} W. Schottky, Phys. Rev. {\bf 28}, 74 (1926).
\bibitem{3} B. B. Mandelbrot, {\it Multifractals and $1/f$ Noise}, (Springer, New York, 1999).
\bibitem{4} P. Dutta and P. M. Horn, Rev. Mod. Phys. {\bf 53}, 497 (1981).
\bibitem{5} P. DeLosRios and Y. C. Zhang, Phys. Rev. Lett. {\bf 82}, 472 (1999).
\bibitem{6} I. Eliazar, and J. Klafter, Phys. Rev. E {\bf 82}, 021109 (2010).
\bibitem{31} P. Helander, S. C. Chapman, R. O. Dendy, G. Rowlands, and N. W. Watkins, Phys. Rev. E {\bf 59}, 6356 (1999).
\bibitem{32} J. Nagler and J. C. Claussen, Phys. Rev. E {\bf 71}, 067103 (2005).
\bibitem{33} J. P. Gleeson, Phys. Rev. E {\bf 72}, 011106 (2005).
\bibitem{34} S. Papanikolau, F. Bohn, R. L. Sommer, G. Durin, S. Zapperi, and J. P. Sethna, Nature Phys. {\bf 7}, 316 (2011).
\bibitem{35} J. P. Sethna, K. A Dahmen, and C. R. Myers, Nature  {\bf 410}, 242 (2001).
\bibitem{38} S. B. Lowen and M. C. Teich, Phys. Rev. E {\bf 47}, 992 (1993).
\bibitem{40} S. Erland and P. E. Greenwood, Phys. Rev. E {\bf 76}, 031114 (2007).
\bibitem{37} J. Ruseckas and B. Kaulakys, Phys. Rev. E {\bf 81}, 031105 (2010).
\bibitem{39} E. W. Montroll and M. F. Shlesinger, Proc. Natl. Acad. Sci. U. S. A. {\bf 79}, 3380 (1982); B. J. West and M. F. Shlesinger, Int. J. Mod. Phys. B {\bf 3}, 795b (1989).
\bibitem{7} P. Bak, {\it How Nature Works: The science of self--organized criticality}, (Copernicus Press, New York, 1996).
\bibitem{8} P. Bak, C. Tang, and K. Wiesenfeld, Phys. Rev. Lett. {\bf 59}, 381 (1987); Phys. Rev. {\bf A 38}, 364 (1988).
\bibitem{9} S. Lubeck and K. D. Usadel, Phys. Rev. E {\bf 56}, 5138 (1997).
\bibitem{10} A. Chessa, E. Marinari, A. Vespignani, and S. Zapperi, Phys. Rev. E {\bf 57}, R6241 (1998).
\bibitem{11} P. DeLosRios, M. Marsili, and M. Vendruscolo, Phys. Rev. Lett. {\bf 80}, 5746 (1998).
\bibitem{laurson} L. Laurson, M. J. Alava, and S. Zapperi, J. Stat. Mech. L11001 (2005).
\bibitem{13} K. Christensen, Z. Olami, and P. Bak, Phys. Rev. Lett. {\bf 68}, 2417 (1992).
\bibitem{Redig} F. Redig, Les Houches Lecture notes (2005).
\bibitem{DD} D. Dhar, Physica A {\bf 369}, 29 (2006).
\bibitem{HJ} H. J. Jensen, {\it Self Organized Criticality} (Cambridge University Press, Cambridge, 1998).
\bibitem{ali} A. A. Ali, Phys. Rev. E {\bf 52}, R4595 (1995).
\bibitem{hwa} T. Hwa and M. Kardar, Phys. Rev. A {\bf 45}, 7002 (1992).
\bibitem{14} S. Maslov, C. Tang, and Y. C. Zhang, Phys. Rev. Lett. {\bf 83}, 2449 (1999).
\bibitem{15} D. Dhar and R. Ramaswamy, Phys. Rev. Lett. {\bf 63}, 1659 (1989).
\bibitem{footer} This  is not really necessary, as the model has the abelian property.
\bibitem{16} D. Dhar, Phys. Rev. Lett. {\bf 64}, 1613 (1990).
\bibitem{pradhan} P. Pradhan and D. Dhar, Phys. Rev. E {\bf 73}, 021303 (2006).

\end{thebibliography}
\end{document}